\documentstyle[11pt,paspconf,epsf]{article}

\newcommand{\Zsun}{Z_{\odot}}

\newcommand{\average}[1]{\langle\rm{#1}\rangle_{\rm V}}

\newcommand{\etal}{{et~al.}}

\newcommand{\tausf}{\tau_{\rm SF}}

\newcommand{\Ae}{$\alpha$-enhancement}

\begin{document}

\title{Star formation histories in early-type galaxies}
\author{D.\ Thomas}
\affil{Universit\"ats-Sternwarte, Scheinerstr.\ 1, 81679 M\"unchen, Germany}


\begin{abstract}
I discuss the formation of $\alpha$-enhanced metal-rich stellar populations
in the nuclei of luminous ellipticals. Based on hierarchical clustering,
different galaxy formation scenarios, which imply different star formation
histories, are considered. In contrast to the {\em fast clumpy collapse}
mode, the late {\em merger of two spiral galaxies} fails to reproduce
significantly $\alpha$-enhanced abundance ratios, unless the IMF is
flattened. Following the star formation history predicted by {\em
semi-analytic models of hierarchical clustering} for the average elliptical,
solar abundance ratios are obtained with Salpeter IMF. According to the
models, bright ellipticals in the field are expected to have significantly
lower Mg/Fe ratios than their counterparts in a cluster.
\end{abstract}



\section{Introduction}
\label{intro}
Theoretical population synthesis models based on {\em solar} abundance
ratios predict -- for a given Fe index -- stronger Mg indices than measured
in bright elliptical galaxies (e.g.\ Worthey, Faber \& Gonz\'{a}lez 1992,
Davies, Sadler \& Peletier 1993). This result is
generally interpreted such that the stellar populations hosted by these
objects are $\alpha$-enhanced with respect to solar values.

\smallskip
In this paper I discuss different formation scenarios for early-type
galaxies and seek under which conditions the observed [Mg/Fe] overabundance
can be achieved.

1) A {\em fast clumpy collapse} is characterized by short star formation
time-scales and implies that ellipticals have formed early at high
redshifts. This picture is observationally supported by the tightness of the
fundamental plane relations (Dressler \etal\ 1987, Renzini \& Ciotti
1993). Also the modest color evolution (Arag\'on
Salamanca \etal\ 1993) and the evolution of the Mg-$\sigma$
relation with redshift (Bender, Ziegler \& Bruzual 1996, Ziegler \&
Bender) push the formation ages of cluster ellipticals to
high redshift ($z>2$). A clumpy collapse in the framework of hierarchical
structure formation can also be considered as a refinement of the classical
monolithic collapse (e.g.\ Matteucci 1994).

2) The scenario in which an elliptical is the outcome of two
{\em merging spirals}, instead, emphasizes the relevancy of the merger
process. Indeed, there are observational indications that merging
must play an important role for the formation of early-type galaxies, since
roughly 50 per cent of bright ellipticals host kinematically decoupled cores
(Bender 1996).

3) The above models can be considered as two extreme cases of substantially
distinct formation scenarios. As a third option I discuss semi-analytic
models of hierarchical clustering, which describe galaxy formation in a
cosmological context (Kauffmann, White \& Guiderdoni 1993, Cole \etal\
1994). A priori, in a bottom-up scenario the star
formation history of a large object is expected to be rather extended.
Hence, the question arises if these kind of models are compatible with
[Mg/Fe] overabundance in luminous ellipticals (Bender 1997).
This issue is investigated here.


\section{The chemical model}
\label{chemical}
The basic constraints on the modeling, aimed to describe the stellar
populations in the central parts of bright ellipticals, are: (1) Achieving
high (super-solar) total metallicities ($Z>\Zsun$), (2) producing an \Ae\ of
the order of [$\alpha$/Fe]$\sim$ 0.2--0.4 dex, as implied by the
observations. The first requirement comes from the observed nuclear Mg$_2$
index and points towards either higher yields (i.e.\ shallow IMF) in the
central regions or to a scenario of {\em enriched} inflow (Edmunds 1990,
Greggio 1997). The latter implies that the composite
stellar populations (CSPs) inhabiting the nuclear regions of ellipticals are
characterized by a metallicity distribution with a minimum metallicity
$Z_{\rm m}>0$. For the {\em fast clumpy collapse} and the {\em merging
spirals} I will show the results for $Z_{\rm m}=\Zsun$. The second
requirement constraints short time-scales or flatter IMF slopes, both
possibilities shall be discussed here. The suggestions of lowering the SN~Ia
rate with respect to our Galaxy and selective mass loss mechanisms, instead,
are not explicitly considered in this paper.

In the models of the sections~\ref{fcc} and~\ref{ms} a normalized amount of
gas is used up and transformed into stars within the star formation
time-scale $\tausf$. The rate of star formation is kept constant. For
details of the modeling I refer the reader to Thomas, Greggio \& Bender
(1998a, 1998b).


\subsection{Fast clumpy collapse}
\label{fcc}
On rather short time-scales ($\sim 1$ Gyr), massive objects are built up by
merging of smaller entities. Star formation takes place within these
entities, on a larger scale, however, the whole system participates in a
general collapse. 
\begin{figure}[ht]
\plotone{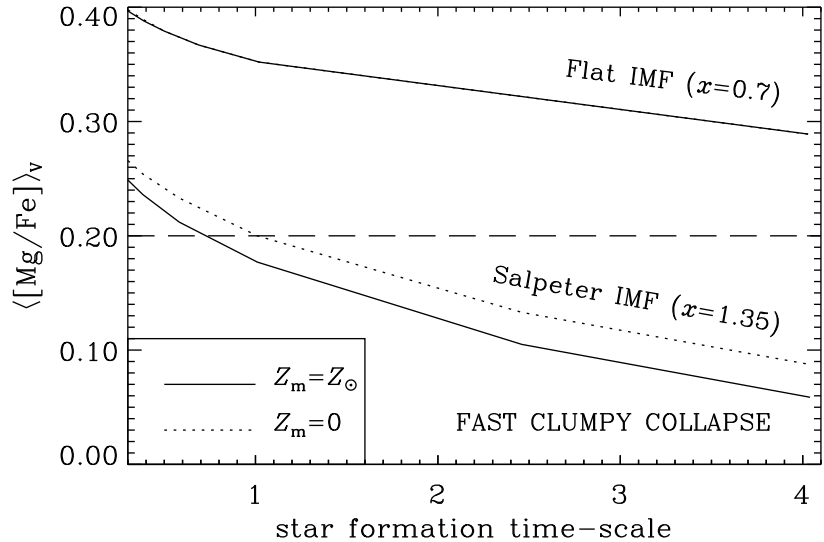}
\caption{$V$-light averaged Mg/Fe ratio in the stars for a family of models
with different star formation time-scales and different IMF slopes.}
\label{fig:fcc}
\end{figure}
Fig.~\ref{fig:fcc} gives the $V$-light averaged Mg/Fe ratio in the stars for
a family of models with different star formation time-scales. The entire
closed-box population with $Z_{\rm m}=0$ (dotted line) and the extracted
metal-rich CSP with $Z_{\rm\-m}=\Zsun$ (solid line) are shown. The result
basically demonstrates that time-scales up to 1~Gyr yield significant \Ae\
for a Salpeter IMF slope ($x=1.35$). The segregation of the metal-rich
population lowers $\average{[Mg/Fe]}$ by less than $\sim\-0.03$ dex, but
increases the metallicity of the CSP by $\sim\-0.4$ dex (not shown in
Fig.~\ref{fig:fcc}, see Thomas \etal\ 1998b). A flattening of
the IMF ($x=0.7$) naturally allows for longer time-scales, and in this case
the cut-off at $Z_{\rm\-m}=\Zsun$ has no significant effect on Mg/Fe.

\subsection{Spiral merger}
\label{ms}
In this scenario I assume that the merger occurs when the two parent spirals
have converted a substantial fraction of gas into stars (i.e.\ two `Milky
Ways'). Hence, most of the stars in the merging entities have formed in a
continuous and long lasting ($\sim$ 10 Gyr) star formation process, leading
to approximately solar abundance ratios. At merging, the (enriched) residual
gas flows down to the center (Barnes \& Hernquist 1996), where
it experiences a violent episode of star formation in which the central
population (considered here) forms on a short burst time-scale. Note that
the global formation process, however, lasts $\sim$ 10 Gyr or more. The
minimum metallicity of the newly formed stars is solar by construction. 
\begin{figure}[ht]
\plotone{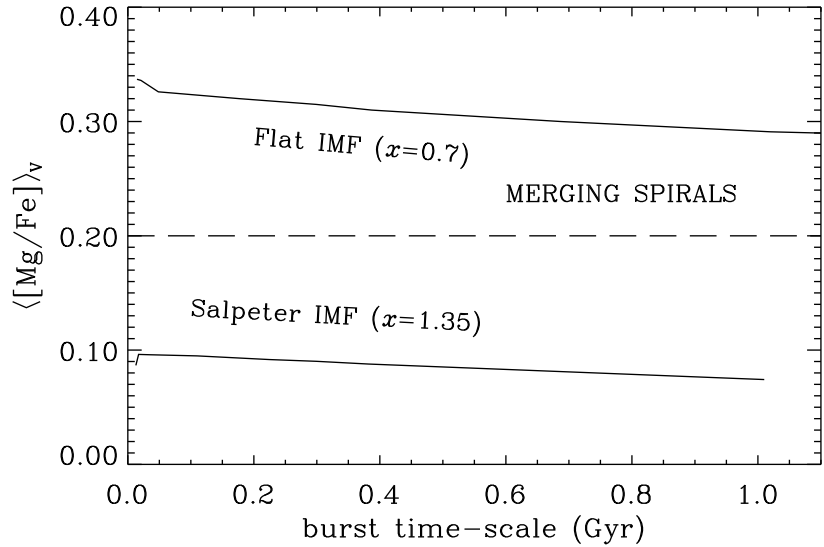}
\caption{$V$-light averaged Mg/Fe ratio in the stars for a family of models
with different burst star formation time-scales and different IMF slopes
{\em during the burst}. Pre-enrichment to solar abundance ratios before the
merger is assumed (merger of two Milky Ways).}
\label{fig:ms}
\end{figure}
For a family of different burst time-scales, the result is shown in
Fig.~\ref{fig:ms} with Salpeter and flat IMF {\em during the burst}. Due to
the huge amount of Fe created in the parent disk galaxy evolution before the
merger, it is now impossible to form significantly $\alpha$-enhanced CSPs
with Salpeter IMF ($x=1.35$), independent of the burst time-scale. In this
case, a flattening of the IMF ($x=0.7$) is required.

\subsection{Hierarchical clustering}
\label{hc}
Semi-analytic models of hierarchical clustering can be placed between the
two extreme cases discussed above. In these simulations structures are
subsequently built up starting from small disk-like objects. An elliptical
is formed when two disk galaxies of comparable mass merge. It is important
to emphasize, that the bulk of stars forms at modest rates during disk
galaxy evolution {\em before} this `major merger' event (Kauffmann 1996,
Baugh, Cole \& Frenk 1996).

To follow the chemical evolution of such models during the entire formation
history, I adopt the star formation history of the {\em average} cluster
elliptical from Kauffmann (1996). Further details are described
in Thomas (1998). 
\begin{figure}[ht]
\plotone{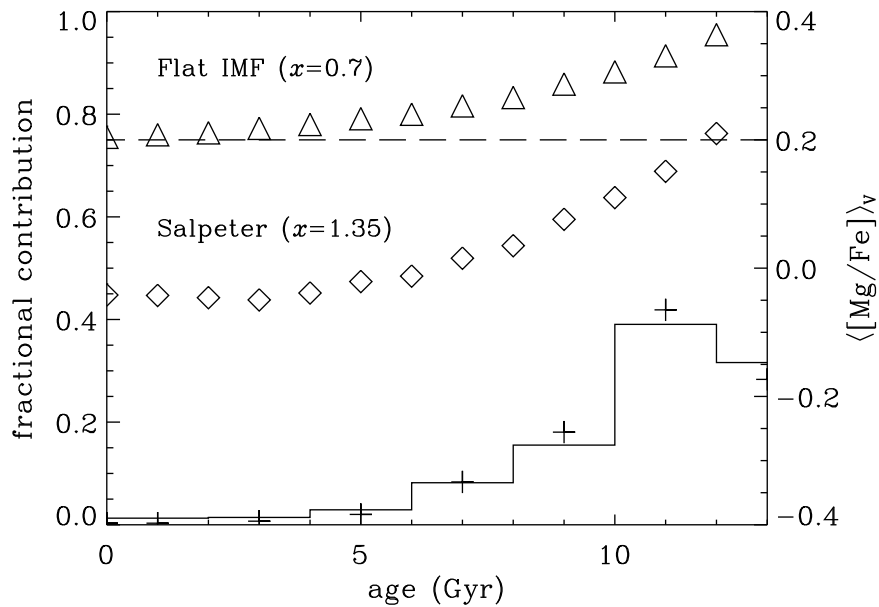}
\caption{The star formation history of an average elliptical galaxy. The
histogram shows the fractional contribution to the total $V$-light (plus
signs: total mass) from each stellar population of a specific age. The
underlying $V$-luminosity averaged abundance ratio Mg/Fe in the stars is
indicated by the diamonds (Salpeter IMF) and triangles (flat IMF).}
\label{fig:k96}
\end{figure}
In Fig.~\ref{fig:k96} the fractional contribution of the stars to the total
$V$-band luminosity of the object is indicated by the solid line (plus signs
show the contribution to the total {\em mass}). The $V$-luminosity weighted
average Mg/Fe ratio of the respective composite stellar population is
indicated by the diamonds for Salpeter IMF ($x=1.35$) and by the triangles
for the flat case ($x=0.7$). Although exhibiting overabundant Mg/Fe ratios
at the early stages of the evolution (i.e.\ the first $1-2$ Gyr),
$\average{[Mg/Fe]}$ decreases with time and saturates at a roughly solar
level after $\sim 10$~Gyr. {\bf\boldmath The star formation history of the
average elliptical as provided by hierarchical models does not lead to \Ae,
unless the IMF is significantly flattened.}

In the Kauffmann (1996) models, the stellar populations of
bright {\em cluster} ellipticals exhibit ages around $11-12$~Gyr, and are
consequently somewhat older than the average elliptical with an age of
$10.5$~Gyr. This implies a star formation history which is more skewed
towards early times and therefore yields higher, probably super-solar Mg/Fe
ratios. Note that, given their short formation times ($1-2$~Gyr), bright
cluster ellipticals form in a scenario which appears similar to the {\em
fast clumpy collapse}. The stellar populations of bright {\em field}
ellipticals, instead, are predicted by the models to be younger by roughly
5~Gyr (Kauffmann 1996). These objects should therefore -- owing
to the more extended star formation history -- exhibit significantly smaller
Mg/Fe ratios than their counterparts in clusters.


\section{Summary}
\label{conclusions}
\begin{description}
\item[(a)] The two distinct formation scenarios, {\em fast clumpy collapse} and 
{\em merging spirals}, yield significantly different abundance ratios.
\item[(b)] In contrast to the fast clumpy collapse scenario, the late merger of two
spirals yields $\alpha$-enhanced populations, only if a shallow IMF is assumed.
\item[(c)] The star formation
history of the average elliptical as specified by {\em hierarchical
clustering} models leads to solar abundance ratios.
\item[(d)] Such models predict significantly
higher Mg/Fe ratios in bright {\em cluster} ellipticals than in bright {\em
field} ellipticals.
\end{description}

\acknowledgments
This work was supported by the "Sonderforschungsbereich 375-95 f\"ur
Astro-Teilchenphysik" of the Deutsche Forschungsgemeinschaft.


\begin{question}{J.\ Gonz\'alez}
Can you comment on our present knowledge of stellar yields? Which ones did
you use and why?
\end{question}
\begin{answer}{D.\ Thomas}
The theoretical yields from Type~II supernovae are still affected by many
uncertainties, particularly magnesium and iron. As a consequence,
conclusions from chemical evolution models are principally weakened.
However, the model and the nucleosynthesis can and should be calibrated on
the chemical evolution of the solar neighborhood. I am using the
yields from Thielemann, Nomoto \& Hashimoto (1996), since they lead to the
best reproduction of the abundance patterns in our Galaxy.
\end{answer}

\begin{question}{R.\ Bower}
In what you discussed, you used a closed box model for chemical evolution.
But in the hierarchical galaxy formation models, the inflow and outflow of
gas are extremely important. How does this affect your conclusions?
\end{question}
\begin{answer}{D.\ Thomas}
In case of hierarchical clustering models, I only discussed abundance {\em
ratios} of elements with no primordial origin. As long as selective loss
mechanisms etc.\ play a minor role, my conclusions on the Mg/Fe ratio are
not affected. But I agree that a more detailed study of chemical enrichment
in the hierarchical models is necessary to enable more reliable quantitative
statements on abundance ratios.
\end{answer}



\begin{references}
\reference Arag{\'o}n~Salamanca, A., Ellis, R.~S., Couch, W.~J.,  \& Carter, D.
  1993, MNRAS, 262, 764
\reference Barnes, J.~E. \& Hernquist, L. 1996, ApJ, 471, 115
\reference Baugh, C.~M., Cole, S.,  \& Frenk, C.~S. 1996, MNRAS, 283, 1361
\reference Bender, R. 1996, in New Light on Galaxy Evolution, R.~Bender \&
  R.~L. Davies, Dordrecht: Kluwer Academic Publishers,  181
\reference Bender, R. 1997, in The second Stromlo Symposium: The nature of
  elliptical galaxies, M.~Arnaboldi, G.~S. Da~Costa, \& P.~Saha, Provo: Brigham
  Young University, 11
\reference Bender, R., Ziegler, B.~L.,  \& Bruzual, G. 1996, ApJ, 463, L51
\reference Cole, S., Arag{\'o}n-Salamanca, A., Frenk, C.~S., Navarro, J.,  \&
  Zepf, S. 1994, MNRAS, 271, 781
\reference Davies, R.~L., Burstein, D., Dressler, A., Faber, S.~M.,
  Lynden-Bell, D., Terlevich, R.~J.,  \& Wegner, G. 1987, ApJS, 64, 581
\reference Davies, R.~L., Sadler, E.~M.,  \& Peletier, R.~F. 1993, MNRAS, 262,
  650
\reference Edmunds, M.~G. 1990, MNRAS, 246, 678
\reference Greggio, L. 1997, MNRAS, 285, 151
\reference Kauffmann, G. 1996, MNRAS, 281, 487
\reference Kauffmann, G., White, S.~D.~M.,  \& Guiderdoni, B. 1993, MNRAS, 264,
  201
\reference Matteucci, F. 1994, A\&A, 288, 57
\reference Renzini, A. \& Ciotti, L. 1993, ApJ, 416, L49
\reference Thomas, D. 1998, MNRAS, submitted
\reference Thomas, D., Greggio, L.,  \& Bender, R. 1998a, MNRAS, 296, 119
\reference Thomas, D., Greggio, L.,  \& Bender, R. 1998b, MNRAS, submitted
\reference Worthey, G., Faber, S.~M.,  \& Gonz{\'{a}}lez, J.~J. 1992, ApJ, 398,
  69
\reference Ziegler, B.~L. \& Bender, R. 1997, MNRAS, 291, 527
\end{references}

\end{document}